\documentclass[aps,prd,twocolumn,superscriptaddress,showpacs]{revtex4}
\usepackage{graphicx}
\usepackage{amsmath}
\usepackage{amsfonts}
\usepackage{amssymb}
\usepackage{amssymb}
\usepackage{slashed}
\usepackage{pstricks,pst-coil}

\begin{document}

\title{Complex-mass shell renormalization of the higher-derivative electrodynamics}

\author{Rodrigo Turcati}\email{rturcati@sissa.it}

\affiliation{SISSA, Via Bonomea 265, 34136 Trieste, Italy and \\
                 INFN, Sezione di Trieste, Via Valerio 2, 34127 Trieste, Italy}
\affiliation{Departamento de F\'isica e Qu\'imica, Universidade Federal do Esp\'irito Santo, Av. Fernando Ferrari, 514, Goiabeiras, 29060-900, Vit\'oria, ES, Brazil}
\affiliation{Laborat\'orio de F\'isica Experimental (LAFEX), Centro Brasileiro de Pesquisas F\'isicas (CBPF), Rua Dr Xavier Sigaud 150, Urca, 22290-180 Rio de Janeiro, Brazil}

\author{Mario Junior Neves}\email{mariojr@ufrrj.br}

\affiliation{Departamento de F\'isica, Universidade Federal Rural do Rio de Janeiro, BR 465-07, Serop\'edica, 23890-971, Rio de Janeiro, Brazil}

\begin{abstract}

We consider a higher-derivative extension of QED modified by the addition of a gauge-invariant dimension-6 kinetic operator in the $U(1)$ gauge sector. The Feynman diagrams at one-loop level are then computed. The modification in the spin-1 sector leads the electron self-energy and vertex corrections diagrams finite in the ultraviolet regime. Indeed, no regularization prescription is used to calculate these diagrams because the modified propagator always occurs coupled to conserved currents. Moreover, besides the usual massless pole in the spin-1 sector, there is the emergence of a massive one, which becomes complex when computing the radiative corrections at one-loop order. This imaginary part defines the finite decay width of the massive mode. To check consistency, we also derive the decay length using the electron--positron elastic scattering and show that both results are equivalent. Because the presence of this unstable mode, the standard renormalization procedures cannot be used and is necessary adopt an appropriate framework to perform the perturbative renormalization. For this purpose, we apply the complex-mass shell scheme (CMS) to renormalize the aforementioned model. As an application of the formalism developed, we estimate a quantum bound on the massive parameter using the measurement of the electron anomalous magnetic moment and compute the Uehling potential. At the end, the renormalization group is analyzed.

\end{abstract}
\maketitle

%
%



\section{Introduction}

Effective field theories (EFT) play a central role in modern physics. They cover almost all branches in physics such as nuclear systems derived from low-energy quantum chromodynamics \cite{kolck}, chiral perturbation theory \cite{weinberg79,gasser1,gasser2,gasser3}, BCS theory formulated from conventional superconductivity \cite{shankar}, inflationary model in cosmology \cite{cheung,senatore}, gravitationally induced decoherence \cite{blencowe}, and so on. Even our most fundamental theories, General Relativity and the Standard Model, are thought of as leading terms of some underlying theory \cite{donoghue1,donoghue2}.

The ideas concerning EFT have began with a nonlinear modification of Maxwell electromagnetism made in order to understand the photon--photon electrodynamical scattering process by Euler and Heisenberg \cite{heisenberg} in the 1930s. At the same time, Fermi developed the theory of beta decay to describe the elementary process $n\rightarrow{p}+e^{-}+\bar{\nu}_{e}$ in the framework of quantum field theory \cite{fermi}. Even having some interesting features, Fermi and Euler--Heisenberg theories were not taken seriously since they were not renormalizable. Nevertheless, some years later, the development of the renormalization and the renormalization group techniques \cite{hooft}, along with the theorem derived by Appelquist and Carazzone \cite{appelquist}---which states that heavy mass particles can be decoupled from low energy dynamics under certain conditions---gave rise to the current EFT programme \cite{weinberg80}.

Effective theories allow us to simplify the description of a given physical process by taking into account the appropriate variables at a given energy scale, i.e., one can consider only the relevant degrees of freedom at a specific energy range. It is basically a low energy dynamics valid below some energy scale and which does not depend on the behavior in the ultraviolet regime. EFT have the advantage of reducing the number of degrees of freedom, turning the description of the physical system under consideration easier to deal with. An appropriate choice of degrees of freedom is a crucial point in the understanding of the problem.

One interesting set of EFT models are the so-called higher-order theories. This class of theories are characterized basically by the modification of the free propagator through the introduction of higher-order kinetic terms in the Lagrangian, which leads to a modified propagator that exhibits a better asymptotic behavior. However, these theories are usually plagued with ghosts states, giving rise to non-Hermitian interactions and scattering processes that do not preserve probabilities, violating the unitarity. There are many phenomena that are described in this framework such as dark energy \cite{Gibbons:2003yj,Carroll:2003st,Woodard:2006nt}, ultraviolet regulators \cite{Slavnov:1972sq,Evens:1990wf,Bakeyev:1996is}, renormalizable gravity models \cite{Stelle:1976gc}, string theory \cite{Eliezer:1989cr}, and supersymmetry \cite{Gama:2011ws}, for instance. It also have been suggested that quantum gravity effects can give rise to partners to every field in the Standard Model \cite{Alvarez:2008za,Wu:2008rr}, which can be included in a straightforward way by the aforementioned formalism, generating an extension of the SM. In addition, some work has recently demonstrated that higher-order derivative theories may emerge in the scope of noncommutative spacetimes \cite{Quesne:2006is,Moayedi:2012fu,Moayedi:2013nba,Moayedi:2013nxa,Silva:2016nmq}. In this framework, the introduction of a minimal length naturally generates higher-order kinetic terms, as in the case of electrodynamics and general relativity \cite{Dias:2016lkg}, for instance. In this sense, a better understanding of some basic features of higher-order theories would certainly help us to improve our knowledge of this set of models.

Concerning QED, a possible way to find an effective Lagrangian is through the addition of a gauge-invariant dimension-6 operator containing higher derivatives in the free Lagrangian of the $U(1)$ sector, namely,
\begin{equation}\label{lweft}
\mathcal{L}_\mathrm{eff}=-\frac{1}{4} \, F_{\mu\nu}^{2}-\frac{1}{4M^2} \, F_{\mu\nu}\Box F^{\mu\nu},
\end{equation}
where $M$ is the only parameter added to theory and is responsible for introducing a cutoff in the ultraviolet regime of QED. This modification is very similar to Pauli--Villars regularization procedure \cite{pauli}, Lee--Wick model \cite{lw69,lw70} at quantum level, and Podolsky electromagnetism \cite{podolsky42,podolsky44} in the classical context. The new degree of freedom in QED changes dramatically the behavior of the theory at short distances. Nevertheless, unitarity is not preserved. \footnote{Actually, the Lee--Wick electrodynamics is claimed to be unitary. Higher-order kinetic terms introduce the appearance of negative norm states in the Hilbert space, which violates the unitarity. Lee and Wick argued that unitarity could be preserved if the Lee--Wick particles obtain a decay width and do not appear in the asymptotic states. For further information, see \cite{lw69,lw70,cut69,coleman70}.} Many remarkable features are found in this QED extension such as the coexistence of Dirac magnetic monopoles and massive vector bosons \cite{turcati14},  the presence of finite electromagnetic mass and self-energy of point particles \cite{podolsky42,podolsky44}. Also, it provides an adequate scenario to solve the 4/3 problem \cite{frenkel96,frenkel99,accioly11}

Our goal in this paper is precisely to address: (i) the one-loop radiative corrections, (ii) the perturbative renormalization of the HDQED in the complex-mass shell (CMS) scheme, (iii) a quantum bound on the $M$-parameter using the measurement of the electron anomalous magnetic moment, (iv) the decay width of the unstable mode taking into account the electron--positron scattering, (iv) the computation of the Uehling potential in the HDQED framework, and the (v) analysis of the renormalization group.

This work is organized as follows. In Sect. 2 we give a brief description of our model. In the following section, Sect. 3, we compute the second-order radiative corrections in the HDQED context. Also, the finite decay width of the unstable mode is explicity calculated. At the end of Sect. 3 we present the Uehling potential. The complex-mass shell renormalization scheme together with the renormalization group is presented in Sect. 4. We summarize our results in Sect. 5.

In our conventions $\hbar=c=1$ and the signature of the metric is ($+1,-1,-1,-1$).


\section{The higher-derivative QED}

The higher-derivative electromagnetism usually is defined by the $U(1)$-gauge-invariant Lagrangian density
\begin{equation} \label{lwl}
\mathcal{L}_\mathrm{eff}=-\frac{1}{4} \, F_{\mu\nu}^{2}-\frac{1}{4M^{2}} \, F_{\mu\nu}\Box{F^{\mu\nu}}-J_{\mu}A^{\mu},
\end{equation}
where $F_{\mu\nu}=\partial_{\mu}A_{\nu}-\partial_{\nu}A_{\mu}$ is the field strength and $M$ is a mass parameter which introduces a length scale in the model. Using the Bianchi identity, the higher-order kinetic term can be rewritten as
\begin{equation}
-\frac{1}{4M^{2}}F^{\mu\nu}\Box{F_{\mu\nu}}\rightarrow\frac{1}{2M^{2}}\partial_{\mu}F^{\mu\nu}\partial^{\lambda}F_{\lambda\nu},
\end{equation}
which implies that Lagrangian (\ref{lwl}) takes the form
\begin{equation} \label{hdqedlagrangian}
\mathcal{L}_\mathrm{eff}=-\frac{1}{4} \, F_{\mu\nu}^{2}+\frac{1}{2M^{2}}\partial_{\mu}F^{\mu\nu}\partial^{\lambda}F_{\lambda\nu}-J_{\mu}A^{\mu}.
\end{equation}

The effective Lagrangian (\ref{hdqedlagrangian}), besides being Lorentz invariant, gives origin to local field equations that are linear in the field quantities and are given by
\begin{eqnarray}
\left(1+\frac{\Box}{M^{2}}\right)\partial_{\mu}F^{\mu\nu}&=&J^{\nu},
\\
\partial_{\mu}\tilde{F}^{\mu\nu}&=&0,
\end{eqnarray}
where $\tilde{F}^{\mu\nu}=\frac{1}{2}\epsilon^{\mu\nu\alpha\beta}F_{\alpha\beta}$. From the above fourth-order equations of motion, it is clear that the system will carry more degrees of freedom than Maxwell electromagnetism. In this sense, it is instructive understand what is the particle content of this model at the quantum level. To accomplish this, we first add the gauge-fixing term $\mathcal{L}_{\zeta}=-\frac{1}{2\zeta}\left(\partial_{\mu}A^{\mu}\right)^{2}$ to the Lagrangian (\ref{hdqedlagrangian}), where $\zeta$ is the gauge-fixing parameter, and then compute the propagator in momentum space, which assumes the form
\begin{equation}\label{lwp}
D_{\mu\nu}(k)=\frac{M^{2}}{k^{2}(k^{2}-M^{2})}\left[\eta_{\mu\nu}+\left(\zeta-1\right) \, \frac{k_{\mu}k_{\nu}}{k^{2}}-\zeta\,\frac{k_{\mu}k_{\nu}}{M^{2}} \right].
\end{equation}

Contracting (\ref{lwp}) with the external conserved currents $J^{\mu}$, one obtains
\begin{equation}\label{amp}
\mathcal{M}\equiv{J^{\mu}(k)D_{\mu\nu}(k)J^{\nu}}(k)=-\frac{J^{2}}{k^{2}}+\frac{J^{2}}{k^{2}-M^{2}}.
\end{equation}

The scattering amplitude ({\ref{amp}}) has poles at $k^{2}=0$ and $k^{2}=M^{2}$. Taking into account that $J$ is space-like ($J^{2}<0$) \cite{accioly03,accioly04,accioly05}, we have the residues
\begin{equation}\label{res}
\mbox{Res}\mathcal{M}(k^{2}=0)>0, \quad \mbox{Res}\mathcal{M}(k^{2}=M^{2})<0.
\end{equation}

From the above residues it is clear that the model under consideration carry two spin-1 modes, one massless and one massive. The positive sign residue corresponds to the spin-1 massless excitation, which can be related to the QED photon. The spin-1 massive mode possess a wrong sign residue which give rise to negative norm states (ghost states) and consequently causes unitarity violation. However, since we are treating the HDQED as an effective field theory, it will not concern us. Regarding to renormalizability, the modified spin-1 propagator always occurs coupled to conserved current, which means that terms proportional to the momenta $k^{\mu}$ give vanish contributions. Hence, the modified propagator reduces to
\begin{equation}\label{lwpesc}
D_{\mu\nu}(k^2)=\frac{M^{2}}{k^{2}(k^{2}-M^{2})} \, \eta_{\mu\nu}=-\frac{\eta_{\mu\nu}}{k^{2}}+\frac{\eta_{\mu\nu}}{k^{2}-M^{2}} \, ,
\end{equation}
which is the difference between Maxwell and Proca propagators. The first term on the right of Eq. (\ref{lwpesc}) is the QED propagator, which is renormalizable by power counting. The second corresponds to a massive spin-1 propagator. Massive vector theories have a bad behavior at high energies and do not go to zero asymptotically. However, there are two exceptions: (i) gauge theories with spontaneous symmetry breaking and (ii) neutral vector bosons coupled to conserved currents. The condition (ii) ensures the renormalizability of the Proca model, showing that the model is renormalizable by power counting.

According to Eq. (\ref{lwpesc}), the nonrelativistic potential $U(r)$ between two electrically charged particle can be expressed as
\begin{eqnarray}\label{pnr}
U(r)&=&U_\mathrm{Coulomb}(r)-U_\mathrm{Yukawa}(r)\\
&=&-\frac{e^2}{4\pi r}\left(1-e^{-Mr}\right),
\end{eqnarray}
which have the following properties:
\begin{enumerate}
\item The regularized potential is finite at the origin
\begin{eqnarray}
U(r\rightarrow 0)=-\frac{e^{2}}{4\pi}M,
\end{eqnarray}
which is evidence that point-like particles has electromagnetic mass and self-energy finities.

\item The standard Coulomb potential is recovered when $Mr>>1$.

\item At short distances, i.e., $Mr<<1$, $U$ differs significantly from the Coulomb potential, as one can see from Fig. 1.
\end{enumerate}

\begin{figure}[h]
\begin{center}
\includegraphics[scale=0.38]{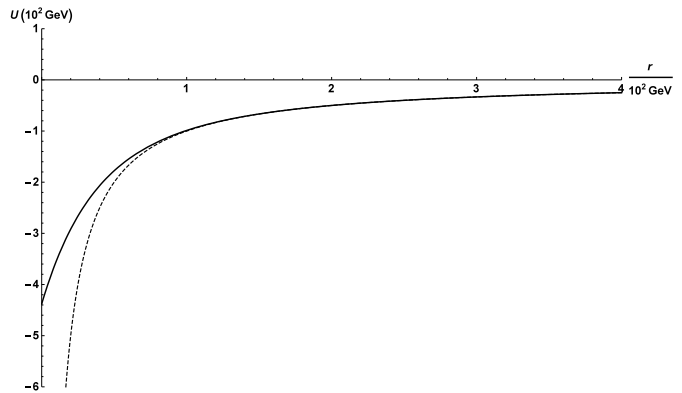}
\end{center}
\vspace{-0.5cm}
\caption{\scshape{The potential $U$ (in units of $\frac{e^2}{4 \pi}$), as a function of the distance $r$.
The dashed line represents the Coulomb potential (in units of $\frac{e^2}{4\pi}$). It is clear from the above figure that in the HDQED the potential is finite at origin. Here, we are assuming the $M$-value found ($M\approx438$ GeV) in Sect. 3.}}\label{potfig}
\end{figure}


\section{Second-order radiative corrections of the HDQED}

The Lagrangian density characterizing the higher-derivative spinor quantum electrodynamics can be written as
\begin{eqnarray}
\mathcal{L}&=&\bar{\psi}\left(i\gamma^{\mu}\partial_{\mu}-m\right)\psi-\frac{1}{4} \, F_{\mu\nu}^{2}
+\frac{1}{2M^{2}} \, \partial_{\mu}F^{\mu\nu}\partial^{\lambda}F_{\lambda\nu}\nonumber\\
&&-e \, \bar{\psi}\gamma^{\mu}\psi{A_{\mu}},
\end{eqnarray}
where the first term on the right is the Dirac kinetic operator, $m$ is the electron mass and the last term denotes the interacting term. In the lowest-order perturbation theory, the scattering of charged particles by each other is characterized by the matrix element of the operator in momentum space
\begin{eqnarray}\label{cc}
\int{\mathrm{d}^{4}k}J^{\mu}(k)D_{\mu\nu}(k)J^{\nu}(k),
\end{eqnarray}
where $J^{\mu}=e\bar{\psi}\gamma^{\mu}\psi$ is the conserved charged fermionic current and $D_{\mu\nu}(k)$ is the Feynman propagator.

According to (\ref{cc}), the modified propagator always occurs coupled to conserved currents and the momenta dependence vanishes, reducing the propagator to
\begin{equation}
D_{\mu\nu}(k)=-\frac{\eta_{\mu\nu}}{k^{2}}+\frac{\eta_{\mu\nu}}{k^{2}-M^{2}}=\frac{M^{2}}{k^{2}(k^{2}-M^{2})} \, \eta_{\mu\nu},
\end{equation}
which provides a better behavior at high energies since the propagator goes as $k^{-4}$ at the asymptotic limit. By power counting, all higher-order process of the Lagrangian (\ref{lwl}) are finite, except for the charge renormalization. In other words, the QED primary divergences related to the electron self-energy and vertex corrections are convergent. Nevertheless, the vacuum polarization diagram, as in QED case, is ultraviolet divergent. We will apply the dimensional regularization to cancel off the divergence arising from the vacuum polarization. Now, we shall undertake the computation of second-order radiative corrections of the HDQED.


\subsection{Electron self-energy}

The electron self-energy expression $\Sigma(p)$ at one-loop approximation in the HDQED is given by
\begin{eqnarray}\label{ese}
-i\Sigma(\slashed{p})&=&(ie)^{2}\int\frac{d^{4}k}{(2\pi)^{4}}\left[\frac{1}{k^{2}-\lambda^{2}}-\frac{1}{k^{2}-M^{2}}\right]\nonumber\\
&&\times\gamma_{\mu} \, \frac{\slashed{p}-\slashed{k}+m}{(p-k)^{2}-m^{2}} \, \gamma^{\mu},
\end{eqnarray}
where $\lambda$ is an infrared regulator. A dimensional analysis shows that the modified propagator makes the integral (\ref{ese}) convergent at high energies. The explicit form of the electron self-energy integral is
\begin{eqnarray}\label{se}
\Sigma(\slashed{p})&=&\frac{\alpha}{2\pi}\int^{1}_{0}dx\left[2m-(1-x)\slashed{p}\right]\nonumber\\
&&\hspace{-0.7cm}\times\mbox{ln}\left[\frac{x(x-1)p^{2}+xm^{2}-(1-x)\mu^{2}+(1-x)M^{2}}{x(x-1)p^{2}+xm^{2}-(1-x)\lambda^{2}}\right],
\nonumber\\
\end{eqnarray}
where $\alpha=e^{2}/4\pi$ is the fine structure constant. Using the on-shell condition ($p'^{2}=p^{2}=m^{2}$) and remembering that the higher-order electrodynamics bring about small deviations from QED, which implies that the mass is very large compared to the electron mass ($M \gg m$), the integral (\ref{se}) reduces to
\begin{eqnarray}
\Sigma\left(\slashed{p}=m\right)
&\cong&\frac{3\alpha}{2\pi}\,m \,\mbox{ln}\left(\frac{M}{m}\right),
\end{eqnarray}
which yields a finite value to the electron mass. Indeed, this result is not a surprise since the self-force acting on a point charge particle is finite and well defined in the HDQED scenario. As an aside, we remark that the 4/3 problem finds a natural explanation in the HDQED context since the electromagnetic mass enters in the Aharonov--Lorentz equation in a form consistent with special relativity \cite{frenkel96,frenkel99,accioly11}. Nevertheless, the infrared divergence remains.


\subsection{Vertex correction}

In the HDQED, the second-order vertex correction is given by
\begin{eqnarray}\label{vc}
\Lambda^{\mu}(p',p)&=&ie^{2}\int\frac{d^{4}k}{(2\pi)^{4}}\frac{M^{2}}{k^{2}\left(k^{2}-M^{2}\right)} \nonumber\\
&&\times\gamma_{\alpha}\frac{(\slashed{p}'-\slashed{k}+m)}{(p'-k)^{2}-m^{2}}\gamma^{\mu}\frac{(\slashed{p}-\slashed{k}+m)}{(p-k)^{2}-m^{2}}\gamma^{\alpha}.\nonumber\\
\end{eqnarray}

As in the case of electron self-energy, the vertex correction (\ref{vc}) is ultraviolet convergent at short distances due to the modified propagator. Again, no regularization procedure is necessary. In order to obtain an explicit form of the vertex correction, we can use the Gordon identity of the Dirac current, which yields
\begin{equation}
\Lambda^{\mu}(p',p)=\gamma^{\mu}F_{1}(q^{2})+\frac{i\sigma^{\mu\nu}q_{\nu}}{2m}F_{2}(q^{2}),
\end{equation}
where the functions $F_{1}(q^{2})$ and $F_{2}(q^{2})$ are called the form factors and $q^{\nu}:=p'^{\nu}-p^{\nu}$ is the four-momentum transfer at the vertex. At radiative corrections, the information concerning to vertex corrections are all contained in $F_{1}(q^{2})$ and $F_{2}(q^{2})$. So, computing the vertex correction at one-loop level provide us with the following form factors:
\begin{eqnarray}
&&F_{1}(q^{2})=\frac{\alpha}{2\pi}\int^{1}_{0}\mathrm{d}x\int^{1}_{0}\mathrm{d}y\int^{1}_{0}dz\delta(1-x-y-z)\nonumber\\
&&\quad\times\left\{ \mbox{ln}\left[\frac{(1-z)^{2}m^{2}-xyq^{2}+zM^{2}}{(1-z)^{2}m^{2}-xyq^{2}}\right]\right.\nonumber\\
&&\quad\left.+\frac{z\,\left[m^{2}(1-4z+z^{2})+(1-x)(1-y)q^{2}\right]}{\left[(1-z)^{2}m^{2}-xyq^{2}\right]
\left[z+(1-z)^{2}\frac{m^{2}}{M^{2}}-xy\frac{q^{2}}{M^{2}}\right]} \right \},\nonumber\\
&&\\
&&\label{f2}
F_{2}(q^{2})=\frac{\alpha}{\pi}\int^{1}_{0}\mathrm{d}x\int^{1}_{0}dy\int^{1}_{0}\mathrm{d}z \,\, z^{2}(1-z)\nonumber\\
&&\quad\times\frac{\delta(1-x-y-z)}{\left[(1-z)^{2}-xy\frac{q^{2}}{m^{2}}\right]\left[z+(1-z)^{2}\frac{m^{2}}{M^{2}}-xy\frac{q^{2}}{M^{2}}\right]}.
\end{eqnarray}

From $F_{1}(q^{2})$ and $F_{2}(q^{2})$ it is clear that ultraviolet divergences do not affect the vertex corrections. The $F_{2}(q^{2})$ form factor when $q^{2} \rightarrow 0$ is related to deviation from the electron magnetic moment standard prediction of the Dirac equation. 


\subsubsection{Anomalous magnetic moment of the electron}

The electron anomalous magnetic moment is the most precise measurement in QED, having an accuracy up to 12 decimal places. This astonishing outcome can be used in the context of HDQED to put a quantum bound on the $M$-parameter and then compare with that of QED.
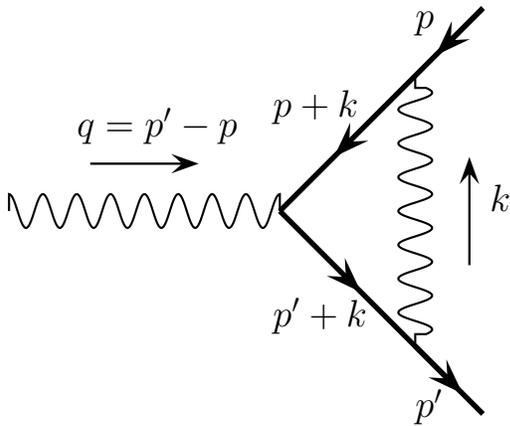
\begin{figure}[!h]
\begin{center}
\newpsobject{showgrid}{psgrid}{subgriddiv=1,griddots=10,gridlabels=6pt}
\begin{pspicture}(0,-0.7)(8,4.5)
\psset{arrowsize=0.2 2}
\psset{unit=0.9}
%
%
%
\pscoil[coilarm=0,coilaspect=0,coilwidth=0.5,coilheight=1.0,linecolor=black](1,2)(5,2)
\psline[linecolor=black,linewidth=0.7mm]{->}(8,5)(7.3,4.3)
\psline[linecolor=black,linewidth=0.7mm]{->}(8,5)(5.8,2.8)
\psline[linecolor=black,linewidth=0.7mm](6,3)(5,2)
\psline[linecolor=black,linewidth=0.7mm]{->}(5,2)(6.2,0.8)
\psline[linecolor=black,linewidth=0.7mm](6,1)(8,-1)
\psline[linecolor=black,linewidth=0.7mm]{->}(7,0)(7.7,-0.7)
\pscoil[coilarm=0.18,coilaspect=0,coilwidth=0.5,coilheight=1.0,linecolor=black](7,0)(7,4)
%
%
%
%
\psline[linecolor=black,linewidth=0.3mm]{->}(2.2,2.7)(3.8,2.7)
\put(2,3.1){\Large$q=p^{\prime}-p$}
\psline[linecolor=black,linewidth=0.3mm]{->}(7.8,1.2)(7.8,2.8)
\put(8.1,2){\Large$k$}
\put(7,4.7){\Large$p$}
\put(7,-1.05){\Large$p^{\prime}$}
\put(4.9,3.4){\Large$p+k$}
\put(4.9,0.3){\Large$p^{\prime}+k$}
%
%
%
\end{pspicture}
\vspace{0.2cm}
\caption{\scshape{The one-loop correction of the vertex diagram.}}\label{figvm}
\end{center}
\end{figure}
To accomplish this, we first note that for an electron scattered by an external static magnetic field and at the limit $ q^{2} \rightarrow 0$, the gyromagnetic ratio is \cite{frampton}
\begin{equation}\label{eamm}
\frac{g}{2}=1+2 \, F_{2}(0) \, .
\end{equation}

The form factor of the electron, $F_2(0)$, corresponds to a shift in the $g$-factor, usually quoted in the form $F_2(0)= \frac{g-2}{2}$. So, taking the limit $q^2 \rightarrow 0$ in (\ref{f2}), it can be shown that
%
\begin{eqnarray}
F_{2}(0)&=&\frac{\alpha}{\pi}\int^{1}_{0}\mathrm{d}x\int^{1}_{0}\mathrm{d}y
\int^{1}_{0}\mathrm{d}z \, \delta(1-x-y-z)\nonumber\\
&&\times\frac{z^{2}}{(1-z)\left[z+(1-z)^{2}\frac{m^{2}}{M^{2}}\right]}.
\end{eqnarray}
Integrating the above expression first with respect to $x$, gives
\begin{eqnarray}
F_{2}(0)&=&\frac{\alpha}{\pi} \int^{1}_{0}\mathrm{d}z \int^{1-z}_{0}\mathrm{d}y \,\, \frac{z^{2}}{(1-z)\left[z+(1-z)^{2}\frac{m^{2}}{M^{2}}\right]}\nonumber\\
&=&\frac{\alpha}{\pi}\int^{1}_{0}\mathrm{d}z \, \, \frac{z^{2}}{z+(1-z)^{2}\frac{m^{2}}{M^{2}}}.
\end{eqnarray}
Computing $F_{2}(0)$, we arrive at the conclusion that
\begin{eqnarray}\label{amm}
F_{2}(0)&\approx&\frac{\alpha}{2\pi}\left\lbrace1-\frac{2}{3}\left(\frac{m}{M}\right)^2 - 2\left[\frac{25}{12}+\ln\left(\frac{m}{M}\right) \right]\left(\frac{m}{M}\right)^4
\right.\nonumber\\
&&+\left.{\cal O}\left[\left(\frac{m}{M}\right)^6\right]\right\rbrace.
\end{eqnarray}

The first term of the above equation is equal to that calculated by Schwinger in 1948 \cite{schwinger48}. Since then  $F_2(0)$ has been calculated to order $\alpha^{10}$ for QED. The second term of Eq. (\ref{amm}) is the most important correction related to the parameter $M$ of the HDQED electrodynamics. Recent calculation concerning $F_2(0)$ in the framework of QED gives for the electron \cite{aoyama12}
\begin{equation}
F_2(0) = 1{,}159{,}652{,}181.643\,(25)(23)(16)(763)\times 10^{-12},
\end{equation}
where the uncertainty comes mostly from that of the best non-QED value of the  fine structure constant $\alpha$. The current experimental value for the anomalous magnetic moment is, in turn \cite{hanneke08,hanneke11}
\begin{equation}
F_2(0) = 1{,}159{,}652{,}180.73\,(0.28)\times 10^{-12}.
\end{equation}

Comparison of the theoretical value predicted by QED  with the experimental one shows that these results agree in $1$  part in  $10^{12}$. As a consequence,
\begin{equation}\label{qblw}
\frac{2}{3}\left(\frac{m}{M}\right)^2<0.91(0.82)\times10^{-12}.
\end{equation}

Consequently, a lower limit on the $M$-parameter is $M\approx$ 438 GeV.
%
%
%
\subsection{Vacuum polarization}

As discussed previously, in the HDQED framework the modification introduced by the higher-order kinetic terms improves the behavior of the physical process in the high frequencies regime. As consequence, the above Feynman diagrams become finite at one-loop order. However, because the fermionic sector is unaltered in the HDQED, the vacuum polarization tensor will be identical to the one in QED, i.e.,
\begin{eqnarray}\label{PolVacuoTransv}
\Pi^{\mu\nu}(k^{2})=\Pi(k^{2}) \, (\eta^{\mu\nu}k^{2}-k^{\mu}k^{\nu}),
\end{eqnarray}
where the contribution at one-loop level of the polarization scalar $\Pi(k^{2})$ is
\begin{eqnarray}\label{Polvacuo1loop}
i\Pi_{1}^{\mu\nu}(k)=-(-ie)^2\int\frac{\mathrm{d}^4p}{(2\pi)^4}\mbox{tr}\left(\frac{i\gamma^{\mu}}{\slash\!\!\!p-m}\frac{i\gamma^{\nu}}{\slash\!\!\!p-\slash\!\!\!k-m}\right).\nonumber \hspace{0.4cm}
\end{eqnarray}

Bearing in mind that the propagator always occurs coupled to conserved current, the dressed propagator in HDQED assumes the form
\begin{eqnarray}\label{fullpropagatorgauge}
i\Delta_{\mu\nu}(k^{2})&=&{}\frac{-i\eta_{\mu\nu}}{k^{2}(1-\Pi(k^{2}))}\nonumber\\
&&+\frac{i\eta_{\mu\nu}}{[k^{2}-M^{2}(1-\Pi(k^{2}))](1-\Pi(k^{2}))},
\end{eqnarray}
which clearly reveals that the radiative corrections coming from the fermionic loops give contribution to $M$-parameter.

Performing the integration of (\ref{Polvacuo1loop}) using dimensional regularization we find
\begin{eqnarray}\label{Pi2epsilon}
&\Pi_{1}(k^{2})=&{}-\frac{\alpha}{3\pi} \frac{\mu^{2\varepsilon}}{\varepsilon}+\frac{\alpha}{6\pi}\,(2\gamma+1)-\frac{\alpha}{3\pi}\ln\left(\frac{4\pi\mu^{2}}{m^{2}}\right)\nonumber\\
&&-\frac{5\alpha}{9\pi} \left(1+\frac{12m^2}{5k^{2}}\right)+\frac{\alpha}{3\pi}\left(1+\frac{2m^{2}}{k^{2}}\right)f(k^{2}),\nonumber \\
\end{eqnarray}
where the function $f(k^{2})$ is given by
\begin{eqnarray}\label{functionfk}
f(k^{2})= \left\{
\begin{array}{lll}
2\sqrt{1-\frac{4m^{2}}{k^{2}}}\sinh^{-1} \! \left(\frac{\sqrt{-k^{2}}}{2m} \right)
\hspace{0.1cm} \mbox{if} \hspace{0.1cm} k^{2}<0 , \\
\sqrt{\frac{4m^{2}}{k^{2}}-1}\cot^{-1} \! \left(\sqrt{\frac{4m^{2}}{k^{2}}-1}\right)
\hspace{0.1cm} \mbox{if} \hspace{0.1cm} 0<k^{2}\leq 4m^{2} \, , \hspace{0.1cm}
\nonumber \\
\sqrt{1-\frac{4m^{2}}{k^{2}}}
\left[ \, 2 \cosh^{-1}\! \left(\frac{\sqrt{k^{2}}}{2m} \right)-i\pi \, \right]
\hspace{0.1cm} \mbox{if} \hspace{0.1cm} k^{2}>4m^{2}.
\end{array}
\right.\nonumber \\ 
\hspace{-1.3cm}
\end{eqnarray}

It is important to note that there is an imaginary part which emerges when $k^{2}>4m^{2}$. This complex term is a clear indication that the massive mode develops a finite decay width, an aspect that will be discussed in detail in the next section. Also, the appearance of a complex massive pole implies that one cannot just apply the standard renormalization schemes. The appropriate framework to treat this issue needs to take into account the existence of unstable particles, a point that deserves some considerations and that will be done in Sect. 6.


\subsubsection{Decay width of the unstable mode}

Before proceeding with the renormalization procedure we will discuss the finite decay rate of the unstable mode. To begin with, we first introduce the auxiliary field formalism. This technique has the advantage of eliminate the higher-order kinetic terms through the introduction of auxiliary fields, reducing the Lagrangian to quadratic terms in the fields. So, defining the auxiliary field
\begin{equation}
Z^{\mu}:=-\frac{\partial_{\lambda}F^{\lambda\mu}}{M^{2}},
\end{equation}
inserting in the Lagrangian (\ref{hdqedlagrangian}), and defining $A^{\mu}:=B^{\mu}+Z^{\mu}$, the Lagrangian (\ref{hdqedlagrangian}) assumes the form
\begin{equation}\label{lagrangianreduced}
\mathcal{L}=-\frac{1}{4} \, B_{\mu\nu}^{2}-\left[-\frac{1}{4} \, Z_{\mu\nu}^{ \, 2}+\frac{1}{2}M^{2} \, Z_{\mu}^{\, 2}\right],
\end{equation}
where $B^{\mu}$ and $Z^{\mu}$ are related to the massless and massive fields, respectively. Here we promptly note that the introduction of auxiliary fields provide us a clear interpretation of the distinct modes at different energy levels. According to (\ref{lagrangianreduced}), at low energy, the massless mode dominates over the massive one the description of the system, reproducing the results of QED. However, at high frequencies, the massive mode emerges and gives contributions to the physical observables as the anomalous magnetic moment of the electron, for instance. Thereafter, at short distances the unstable mode should develop a finite decay width. In the auxiliary field formalism the decay width description can be simplified by noting that at the ultraviolet regime only the massive propagator is relevant, namely,
\begin{equation}
D_{\mu\nu}^{M}(k^2)=\frac{\eta_{\mu\nu}}{k^{2}-M^{2}},
\end{equation}
which is the Proca propagator with a minus sign. In electron--positron scattering, when the mass of the massive mode exceeds the electron--positron mass ($M>2m_{e}$), the negative residue appears in the physical sheet. However, as we have discussed in the previous section, at this energy range the massive mode becomes complex and must decay in light particles. The decay width $\Gamma$ can be obtained by the standard self-energy sum. Hence, at the narrow width approximation, the resummed massive propagator assumes the form
\begin{equation}\label{narrowpropagator}
D_{\mu\nu}^{M}(k)=\frac{\eta_{\mu\nu}}{k^{2}-M^{2}-iM\Gamma}.
\end{equation}

Comparing the denominator of (\ref{narrowpropagator}) with the imaginary part that arises in the self-energy sum when $k^{2}>{m}^{2}$ in (\ref{functionfk}), we promptly find that the finite decay width of the $M$-parameter is
\begin{eqnarray}\label{decayrate}
\Gamma=\frac{e^{2}}{12\pi}M\left(1+\frac{2m^{2}}{M^{2}}\right)
\sqrt{1-\frac{4m^{2}}{M^{2}}}.
\end{eqnarray}

To check the consistency of the previous decay width derivation, one can follow another route to obtain the decay rate of the $M$-parameter through the analysis of the electron--positron elastic scattering. The general expression for the decay rate of this process is
\begin{equation}
\Gamma=\frac{1}{2\pi^{2}}\frac{1}{2k^{0}}\int\frac{\mathrm{d}^{3}q}{2q^{0}} \int\frac{\mathrm{d}^{3}q'}{2q'^{0}} \, \delta^{4}(k-q-q') \, \frac{1}{3}\sum_{\lambda,s,s'}|\mathcal{M}|^{2},
\end{equation}
where $k$ is the massive spin-1 four-momentum and $q$ and $q'$ are related to the electron and positron momenta, respectively. The electron--positron elastic scattering amplitude is
\begin{equation}
\mathcal{M}=e\epsilon_{\mu}(k,\lambda)\bar{u}(q,s)\gamma^{\mu}v(q',s'),
\end{equation}
where $\epsilon^{\mu}$ is the polarization vector and $u$ and $v$ represent the fermions functions. Using the completeness relation,
\begin{eqnarray}
\sum_{\lambda}\epsilon_{\mu}(k,\lambda)\epsilon_{\nu}(k,\lambda)=-\eta_{\mu\nu}+\frac{k_{\mu}k_{\nu}}{M^{2}},
\\
\sum_{s}u_{\alpha}(q,s)\bar{u}_{\beta}(q,s)=\left(\slashed{q}+m\right)_{\alpha\beta},
\\
\sum_{s'}v_{\alpha}(q',s')\bar{v}_{\beta}(q',s')=\left(\slashed{q}'-m\right)_{\alpha\beta},
\end{eqnarray}
the decay factor $\Gamma$ assumes the form
\begin{eqnarray}
\Gamma&=&\frac{e^{2}}{3\pi^{2}}\frac{1}{M^{3}}\int\frac{\mathrm{d}^{3}q}{2q^{0}}\int\frac{\mathrm{d}^{3}q'}{2q'^{0}}\delta^{4}(k-q-q')\left[(q\cdot{k})(q'\cdot{k})\right.\nonumber\\
&&\left.+\frac{1}{4}M^{2}(M^{2}+4m^{2})\right].
\end{eqnarray}

Solving the above integral, we arrive at the decay rate
\begin{equation}
\Gamma=\frac{e^{2}}{12\pi}M\sqrt{1-4\frac{m^{2}}{M^{2}}}\left(1+2\frac{m^{2}}{M^{2}}\right) \theta(M-2m),
\end{equation}
which is identical to the decay rate (\ref{decayrate}) found when considering the one-loop corrections in the spin-1 sector, so lending support to our derivation. It is worth to note that since $M\gg2m$, the $\Gamma$ factor reduces to
\begin{equation}
\Gamma\approx\frac{e^{2}}{12\pi}M.
\end{equation}

In general, there may be many decays modes, which means that the massive mode lifetime $\tau$ is given by
\begin{equation}
\tau=\frac{1}{\Gamma_{T}}<\frac{12\pi}{e^{2}M},
\end{equation}
where using our estimative for the mass value $M\approx438$ GeV gives a lifetime of $\tau\lesssim10^{-24}s$.


\subsubsection{Uehling potential}

One of the great triumphs of QED was the prediction of the small difference between the energy levels $^{2}S_{1/2}$ and $^{2}P_{1/2}$ in the hydrogen atom---the Lamb shift---which is related to the radiative corrections to the Coulomb potential coming from the vacuum polarization. Since in the HDQED framework the energy potential is not singular at the origin \cite{accioly11}, we are interested in probe what sort of contribution will arise in this context. To start with, the general expression for the energy potential is given by
\begin{eqnarray}\label{UIntFourier}
U({\bf r})=-4\pi\alpha\int\frac{\mathrm{d}^3{\bf k}}{(2\pi)^{3}}e^{i{\bf k}\cdot{\bf r}}\Delta_{00}\left(-{\bf k}^{2}\right).
\end{eqnarray}

Inserting the resummed propagator (\ref{fullpropagatorgauge}) in the energy potential expression (\ref{UIntFourier}), and neglecting terms of $\alpha^{3}$ order, one obtains
\begin{eqnarray}\label{UintdkSimpZeta}
&&U(r)=-\frac{\alpha}{r} \left(1-e^{-Mr}\right) \nonumber \\
&&\quad-\frac{2\alpha^{2}}{3\pi r}
\int_{1}^{\infty} \!\!\!\! d\xi \left(1+\frac{1}{2\xi^{2}} \right) \frac{\left(\xi^{2}-1 \right)^{1/2}}{\xi^{2}}\frac{e^{-2mr\xi}-e^{-Mr}}{1-4m^{2}\xi^{2}/M^{2}}
\nonumber \\
&&\quad+\frac{2\alpha^{2}}{3\pi r}\frac{\partial}{\partial M^{2}}\int_{1}^{\infty} \!\!\!\! d\xi \left(1+\frac{1}{2\xi^{2}} \right) \frac{\left(\xi^{2}-1 \right)^{1/2}}{\xi^{2}}\frac{e^{-2mr\xi}-e^{-Mr}}{1-4m^{2}\xi^{2}/M^{2}}
\nonumber \\
&&\quad-\frac{\alpha}{2r}Mre^{-Mr}\frac{2\alpha}{\pi} \int_{0}^{1}\mathrm{d}x \, x(1-x)\ln\left[1-\frac{M^{2}}{m^{2}}x(1-x)\right] \, ,
\hspace{-2.5cm}\nonumber \\
\end{eqnarray}
where the $\xi$-integral above is the integral representation of the {\it Uehling potential} in the HDQED framework. Since the above integral is hard to solve analytically, we will focus only on the asymptotic limit $mr \gg 1$.

For $mr \gg 1$, only the region $0 \leq \xi-1 \ll \left(mr\right)^{-1}$ gives contributions to the integral. So, considering $\xi \simeq 1$, we obtain
\begin{eqnarray}\label{UintdkSimpZetaapprox}
&&U(r)=-\frac{\alpha}{r} \left(1-e^{-Mr}\right)\nonumber\\
&&\quad-\sqrt{2} \, \, \frac{\alpha^{2}}{\pi r} \,
\int_{1}^{\infty} \mathrm{d}\xi \, \, \sqrt{\xi-1} \; \; \frac{e^{-2mr\xi}-e^{-Mr}}{1-4m^{2}\xi^{2}/M^{2}}+
\nonumber \\
&&\quad+ \sqrt{2} \, \, \frac{\alpha^{2}}{\pi r} \, \, \frac{\partial}{\partial M^{2}}
\int_{1}^{\infty} d\xi \, \, \sqrt{\xi-1} \; \; \frac{e^{-2mr\xi}-e^{-Mr}}{1-4m^{2}\xi^{2}/M^{2}}+
\nonumber \\
&&\quad+\frac{\alpha^{2}}{2 \pi r}Mr \, \, e^{-Mr} \left\{\frac{5}{9} \left(1+\frac{12m^2}{5M^{2}}\right)\right. \nonumber \\
&&\quad\left.-\frac{1}{3}\left(1+\frac{2m^{2}}{M^{2}}\right)
\sqrt{1-\frac{4m^{2}}{M^{2}}} \,
\left[2 \cosh^{-1} \left(\frac{M}{2m} \right)-i\pi \right] \right\}.\nonumber \\
\end{eqnarray}

In the above limit, the $\xi$-integral becomes
\begin{eqnarray}
&&\int_{1}^{\infty} \mathrm{d}\xi \, \, \sqrt{\xi-1} \; \; \frac{e^{-2mr\xi}-e^{-Mr}}{1-4m^{2}\xi^{2}/M^{2}} \nonumber\\
&&\quad\frac{\sqrt{\pi}}{4}\left(1+\frac{2m}{M}\right)^{1/2}\left(\frac{M}{2m}\right)^{3/2} e^{Mr} \, \, \Gamma\left(-\frac{1}{2},Mr+2mr\right)
\nonumber \\
&&\qquad+\left(\frac{\pi}{2mr}\right)^{1/2} e^{-2mr} \, \frac{M}{4m} \, \Phi\left(1,\frac{3}{2},-Mr+2mr\right) \nonumber \\
&&\qquad+\frac{\pi}{2}\left(1+\frac{2m}{M}\right)^{1/2}\left(\frac{M}{2m} \right)^{3/2}e^{-Mr},
\end{eqnarray}
where $\Gamma$ is the incomplete Gamma function and $\Phi$ is the confluent hypergeometric function of the first kind (Kummer's function).
Since $M>2m$ in (\ref{UintdkSimpZeta}), the Uehling potential also gives rise to an imaginary part, which goes to zero when $M$ goes to infinity. At the limit in which the massive mode is much bigger than the mass of the fermions, i.e., $M \gg m$, the perturbative expansion at $m/M$ gives for the real part of the Uehling potential, at the first order, the expression
\begin{eqnarray}\label{ReU}
&&\Re(U)\simeq-\frac{\alpha}{r}\left(1-e^{-Mr}\right)-\frac{\alpha^{2}}{\sqrt{2\pi}r}\frac{e^{-2mr}}{(2mr)^{3/2}}\left(1-\frac{3}{4Mr} \right)
\nonumber \\
&&- \, \frac{\alpha^{2}}{\sqrt{2} \, r} \left(\frac{M}{2m} \right)^{3/2} \!\! e^{-Mr}
+\frac{11\alpha^{2}}{18\pi} \, M \, e^{-Mr},
\end{eqnarray}
while for the imaginary part it gives
\begin{eqnarray}\label{ImU}
\Im(U) \simeq \frac{\alpha^{2}}{6} \, M \, e^{-Mr}+ \, \, \frac{\alpha^{2}}{4\sqrt{2} \, r}
\, \, \frac{1}{\left(2mr\right)^{3/2}} \, \frac{e^{-Mr}}{\left(Mr\right)^{1/2}},
\end{eqnarray}
which exponentially decay at the asymptotic limit.

To conclude we remark that in the limit that $M$ goes to infinity, the $\xi$-integral becomes a Gamma function and the potential (\ref{UintdkSimpZeta}) reduces to
\begin{eqnarray}\label{UintdkSimpZetalinhaapprox}
\lim_{M\rightarrow \infty}U(r)=-\frac{\alpha}{r}-\frac{\alpha^{2}}{\sqrt{2\pi} \, r}
\, \frac{e^{-2mr}}{(2 m r)^{3/2}},
\end{eqnarray}
which is the standard Uehling potential in QED.

\section{Renormalization}
\subsection{Renormalized perturbation theory}

The presence of a finite decay width implies that one needs to find an appropriate framework to describe unstable particles in the perturbation theory. It follows that applying directly the standard summation of self-energy violates the gauge invariance \cite{Berends69,Kurihara94,Argyres95}. To circumvent this problem, many methods have been proposed over the last years \cite{Passarino:1999zh,Baur:1995aa,Aeppli:1993rs,Stuart:1991xk,Beenakker:1996kn,Aeppli:1993cb,Accomando:1999zq,Beenakker:2003va,Beenakker:1999hi,Beneke:2004km,Beneke:2003xh}. Currently, the most general procedure to treat perturbative renormalization of unstable particles in quantum field theory is the so-called complex-mass shell (CMS) scheme. The CMS is an extension of the on-shell procedure for unstable particles which is fully gauge invariant over all the phase space. In this formalism, the renormalization constant and the complex massive pole are defined at the location of the unstable propagator (for further information, see \cite{Denner99,Denner05,Denner06}).

Performing the perturbative renormalization one obtains the renormalized Lagrangian
\begin{eqnarray}\label{renormalizedlagrangian}
\mathcal{L}&=&\bar{\psi}\left(i\gamma^{\mu}\partial_{\mu}-m\right)\psi-\frac{1}{4}F_{\mu\nu}^{2}
+\frac{1}{2M^{2}}\partial_{\mu}F^{\mu\nu}\partial^{\lambda}F_{\lambda\nu}\nonumber\\
&&-e\bar{\psi}\gamma^{\mu}\psi{A_{\mu}}-\frac{1}{2\zeta}\left(\partial_{\mu}A^{\mu}\right)^{2}+i\delta_{\psi}\bar{\psi}\gamma^{\mu}\partial_{\mu}\psi-\delta_{m}\bar{\psi}\psi\nonumber\\
&&
-\frac{\delta_{A}}{4}F_{\mu\nu}^{2}+\frac{1}{2\delta_{M}}\partial_{\mu}F^{\mu\nu}\partial^{\lambda}F_{\lambda\nu}-\delta_{e}\bar{\psi}\gamma^{\mu}\psi{A_{\mu}},
\end{eqnarray}
where the relations between the bare and renormalized quantities are
\begin{eqnarray}
\psi^{(0)}=\sqrt{Z_{\psi}} \, \psi, \quad A^{(0)}_{\mu}=\sqrt{Z_{A}} \, A_{\mu},
\end{eqnarray}
and
\begin{equation}\label{contratermosQED}
\zeta_{0}=Z_{A} \, \zeta, \quad e_{0}\sqrt{Z_{A}}=e, \quad \frac{Z_{A}}{M_{0}^{2}}=\frac{1}{M^{2}}+\frac{1}{\delta_{M}}.
\end{equation}

We call attention to the fact that, according to the usual renormalization procedure, one has to put counterterms into all diagrams, even the finite ones, since in an interacting theory the bare mass and the coupling constants are not equal to the physical parameters. A quick glance at the renormalized Lagrangian (\ref{renormalizedlagrangian}) shows us that five conditions are necessary in order to fix all the counterterms, which are given by
\begin{eqnarray}\label{CondRenor}
&&\Sigma(\slashed{q})|_{\slashed{q}=m}=0, \quad \frac{d}{d\slashed{q}}\Sigma(\slashed{q})|_{\slashed{q}=m}=0, \quad \Pi(k^{2})|_{k^{2}=0}=0,\nonumber \\
&&\hspace{0.9cm}\Gamma^{\mu}(p)|_{p=0}=\gamma^{\mu}, \quad\quad \Pi(k^{2})|_{k^{2}=M^{2}}=0.
\end{eqnarray}

The first constraint fixes the physical electron mass, the second and third fix the Dirac and sector-1 propagator to have a residue equal to 1, the fourth fixes the electron charge, and the last one defines the complex massive pole. Since the massive pole is complex, the renormalization constant will also be complex, which agrees with our claim that one needs to find a convenient framework to deal with the renormalization of unstable particles.
In this sense, writing $M_{0}=\sqrt{Z_{M}} \, M$, where $Z_{M}$ is the complex massive renormalization factor defined by
\begin{eqnarray}\label{ZMPolo}
Z_{M}=\frac{1}{1-\Pi(k^{2}=M^{2})},
\end{eqnarray}
one obtains the relation $M^{2}/M_{0}^{2}:=1-\Pi\left(k^{2}=M^{2}\right)$.

On the other hand, the $M$-parameter is renormalized by the renormalization function $Z_{A}$. As well as QED, the $Z_{A}$-factor is given by
\begin{eqnarray}\label{ZAPolo}
Z_{A}=\frac{1}{1-\Pi(k^{2}=0)},
\end{eqnarray}
and the renormalization factors $Z_{A}$ and $Z_{M}$ are related by the expression
\begin{eqnarray}\label{ZAZM}
Z_{M}=Z_{A}\,\left(1+\frac{M^{2}}{\delta_{M}} \right)^{-1}.
\end{eqnarray}

We also point out that the Ward identity ensures that the factor $Z_{\psi}$ is equal to the vertex correction renormalization $Z_{V}$, i.e., $Z_{\psi}=Z_{V}$. As a consequence, the HDQED does not give a further contribution to the charge renormalization.


\subsection{Renormalization group}

The Callan--Symanzik equation of the renormalization group is given by
\begin{eqnarray}\label{Callan}
\left[\mu\frac{\partial}{\partial \mu}+\beta(e)\frac{\partial}{\partial e}+M\gamma_{M}(e)\frac{\partial}{\partial M}-n\gamma_{A}(e)\right]\Gamma^{(n)}=0,\nonumber\\
\end{eqnarray}
where $\Gamma^{(n)}$ is the one-particle irreducible Green function and $\mu$ is an arbitrary energy scale. The functions $(\beta, \gamma_{M}, \gamma_{A})$ are related with the renormalization factors by
\begin{eqnarray}\label{betagamma}
&&\beta(e)=\mu\,\frac{\partial e}{\partial \mu}, \quad
\gamma_{M}(e)=-\frac{\mu}{2} \, \frac{\partial}{ \partial \mu}\ln Z_{M}
\quad \mbox{and} \nonumber \\
&&\quad\gamma_{A}(e)=\frac{\mu}{2} \, \frac{\partial}{ \partial \mu}\ln Z_{A}.
\end{eqnarray}

Thus we may conclude that the beta function is unchanged in the HDQED framework. This happens because at high momenta the $M$-parameter is negligible, being given explicitly by $\beta(e)=e^{3}/12\pi^{2}$. Moreover, as well as in QED, the HDQED is not asymptotically free. Combining the functions (\ref{betagamma}) and Eqs. (\ref{contratermosQED}), it is easy to see that $\beta(e)=e\, \gamma_{A}(e)$, implying that $\gamma_{A}(e)=\alpha/3\pi$ and
\begin{eqnarray}\label{gammaM}
\gamma_{M}(e)=-\frac{e^{2}}{12\pi^{2}}.
\end{eqnarray}

The invariance of the Green function $\Gamma^{(n)}$ under a scale transformation, $\Gamma^{(n)}(e,M,\mu)=\Gamma^{(n)}(\bar{e}(t),\bar{M}(t),\bar{\mu}(t)=\mu\,e^{t})$, leads to an {\it effective constant coupling} $\bar{e}(t)$ and an {\it effective $\bar{M}(t)$-parameter} as a function of the dimensionless scale $t$, both satisfying the equations
\begin{eqnarray}\label{eMeffective}
&\frac{\partial \bar{e}}{\partial t}=\beta(\bar{e}),\\
\label{Meffective}
&\frac{\partial \bar{M}}{\partial t}=\bar{M}(t) \, \gamma_{M}(\bar{e}(t)),
\end{eqnarray}
where $\bar{e}(t=0)=e$ and $\bar{M}(t=0)=M$. The solution of (\ref{eMeffective}) gives
\begin{eqnarray}\label{et}
\bar{e}^{2}(t)=e^{2}\left(1-\frac{e^{2}\,t}{6\pi^{2}}\right)^{-1}.
\end{eqnarray}

Solving Eq. (\ref{Meffective}) we find
\begin{eqnarray}\label{LWMassEffective}
\bar{M}(t)= M \, \frac{e}{\bar{e}(t)} =M \, \sqrt{1-\frac{e^{2}\, t}{6\pi^{2}}}.
\end{eqnarray}
which gives a running mass as a function of the energy scale.
%


\section{Final remarks}

In this paper we have investigated some issues related to one-loop corrections of HDQED. Since the higher-order kinetic terms in the spin-1 sector improves the behavior of the QED propagator at short distances, we have shown that the vertex correction and the electron self-energy are finite and depends explicitly on the $M$-parameter. In this sense, it became clear that the $M$-parameter acts as a natural regulator parameter which at the limit where $M\rightarrow\infty$ recovers the QED divergences. Nevertheless, the vacuum polarization remains divergent. It is also worth to emphasize that we did not make use of any regularization procedure in the computations of these diagrams. The only assumption made was that the scattering processes should occur in the presence of external conserved currents. Despite the fact that the couplings with external conserved currents have eliminated the momenta dependence in the loop integrals, which simplified our task of solving the integrals, at the next leading order this prescription cannot be used anymore. Higher-order scattering processes induce the appearance of internal loops, which are not canceled when one takes into account the presence of conserved currents. We also call attention to the fact that our computations were made in the Lorenz gauge. However, any gauge condition is feasible \cite{turcati14}.

From the analysis of the complex-mass shell conditions in the perturbative renormalization scheme, we have concluded that the HDQED enforces us to input another constraint in order to fix all the counterterms in the renormalized Lagrangian. Concerning the charge renormalization, the massless pole is the only one relevant while for the renormalization of the $M$-parameter one needs to take into account the complex massive pole. Analyzing the Callan--Symanzik equation, we have observed that the beta function and the anomalous dimension are unchanged by the presence of the $M$-parameter. Thus we may conclude that the $M$-parameter is negligible at the ultraviolet regime. Nevertheless, the gamma function associated to the $M$-parameter given by (\ref{gammaM}) and the effective $M$-parameter has an interesting behavior running with the energy scale. It is worth to remark that to the best of our knowledge it is the first time that the CMS scheme is applied to treat higher-order derivative electrodynamics theories.

As an interesting application we have found a quantum bound on the $M$-parameter using the value of the electron anomalous magnetic moment. This measurement has been used to estimate stringent constraints on possible theories beyond the standard model. Using the latest measurements of the mentioned phenomenon we found $M=438$ GeV. Also, we have calculated the Uehling potential in the HDQED framework. Although the effect of Uehling potential to the Lamb shift is negligible, in muonic atoms the vacuum polarization is dominant, which in principle could be used to put a new bound in the mass parameter of HDQED. This new constraint on the $M$-parameter can be further investigated in the future. Regarding the modern particle physics experiments, we point out that the detection of this new degree of freedom could be related to physics beyond the Standard Model. In our prescription we have included the higher-order term after the spontaneous symmetry breaking has occurred in the electroweak theory (EW). However, this new degree of freedom could be added in the EW Lagrangian before the symmetry breakdown. In this way, the appearance of a mass$^{-2}$ dimensional parameter at TeV scale could be a sign of a phenomenon arising from new physics, as effects of a large extra dimension, for instance.

In order to outline possible directions for future investigations, we would like to emphasize that the scenario developed here to treat unstable modes could be extended, at least in principle, to other spin fields. Also, issues concerning the unitarity of higher-order theories in the CMS formalism can be investigated in the future.


\section{Acknowledgments}

The authors are grateful to A. J. Accioly, J. A. Helay\"el-Neto and A. D. Pereira for the comments on a draft of this manuscript. Rodrigo Turcati is very grateful to CNPq (Brazilian agency) for financial support.



%
%
%
%
%
%

%
%
\end{document}